\documentclass[onecolumn]{revtex4}

\usepackage{graphicx}
\usepackage{dcolumn}
\usepackage{bm}

\input epsf

\begin{document}

\title{\Large Validity of Generalized Second Law of
Thermodynamics in the \emph{Logamediate} and \emph{Intermediate}
scenarios of the Universe}

\author{\bf Arundhati Das$^1$\footnote{arundhatiwonderland@gmail.com}, Surajit
Chattopadhyay$^2$\footnote{surajit$_{_{-}}2008$@yahoo.co.in,
surajit.chattopadhyay@pcmt-india.net} and Ujjal
Debnath$^1$\footnote{ujjaldebnath@yahoo.com,
ujjal@iucaa.ernet.in}}

\affiliation{$^1$Department of Mathematics, Bengal Engineering and
Science University, Shibpur, Howrah-711 103,
India.\\
$^2$Department of Computer Application (Mathematics Section),
Pailan College of Management and Technology, Bengal Pailan Park,
Kolkata-700 104, India.}

\begin{abstract}
In this work, we have investigated the validity of the generalized
second law of thermodynamics in \emph{logamediate} and
\emph{intermediate} scenarios of the universe bounded by the
Hubble, apparent, particle and event horizons using and without
using first law of thermodynamics. We have observed that the GSL
is valid for Hubble, apparent, particle and event horizons of the
universe in the logamediate scenario of the universe using first
law and without using first law. Similarly the GSL is valid for
all horizons in the intermediate scenario of the universe using
first law. Also in the intermediate scenario of the universe, the
GSL is valid for Hubble, apparent and particle horizons but it
breaks down whenever we consider the universe enveloped by the
event horizon.
\end{abstract}

\maketitle

\section{\normalsize\bf{Introduction}}

Nowadays, it is widely accepted fact that the universe is
experiencing accelerated expansion driven by dark energy [1]
characterized by negative pressure $p_{\Lambda}$ satisfying the
equation of state parameter $w_{\Lambda}<-1/3$ [2]. Strong
observational evidences for dark energy are available from Type Ia
supernova (SN Ia), cosmic microwave background radiation and Sloan
digital sky survey (SDSS) observations [3] that entities in modern
physics. Reviews on dark energy are available in [4].  Although
observationally well-established, no single theoretical model
provides an eternally compelling framework within which cosmic
acceleration or dark energy can provide an entirely compelling
framework within which the dark energy can be well-understood.
Several models for dark energy have been proposed till date. Such
models include quintessence, phantom, quintom, holographic dark
energy etc. Discussions on these models are available in Copeland
et al [4]. Determining thermodynamic parameters for the expanding
(accelerated) universe and verification of the first and the
second law for different cosmological horizons, investigating the
relation between dynamics and thermodynamics of the universe,
studying the conditions required for validity of the generalized
second law (GSL) have also been the subjects of interest in recent
years [5].\\

Since the discovery of black hole thermodynamics in 1970's,
physicists have been speculating that there should be some
relation between black hole thermodynamics and Einstein equations.
In 1995, Jacobson [6] derived Einstein equations by applying the
first law of thermodynamics $\delta Q = T dS$ together with
proportionality of entropy to the horizon area of the black hole.
Here $\delta Q$ and $T$ are the energy flux and Unruh temperature
seen by an accelerated observer just inside the horizon. Verlinde
[7] found that the Friedmann equation in a radiation dominated
Friedmann-Robertson-Walker (FRW) universe can be written in an
analogous form of the Cardy-Verlinde formula, an entropy formula
for a conformal field theory. The first law of thermodynamics for
the cosmological horizon is given by $-dE=TdS$, where
$T=\frac{1}{2\pi l}$ is the Hawking temperature, and
$S=\frac{A}{4G}$ is the entropy with $A=4\pi l^{2}$ and $G$ as the
cosmological horizon area and Newton constant respectively (Cai
and Kim in ref [8]). Einstein's field equations have been derived
from the first law of thermodynamics in the references given in
[8]. In [9], the gravitational field equations for the nonlinear
theory of gravity were derived from the first law of
thermodynamics by adding some non equilibrium corrections.
Profound physical connection between first law of thermodynamics
of the apparent horizon and the Friedmann equation was established
in [10].\\

In a spatially flat de Sitter space–time, the event horizon and
the apparent horizon of the Universe coincide and there is only
one cosmological horizon. When the apparent horizon and the event
horizon of the Universe are different, it was found that the first
law and the second law of thermodynamics hold on the apparent
horizon, while they break down if one considers the event horizon
[11]. There are several studies in thermodynamics for dark energy
filled universe on apparent and event horizons [12]. Setare and
Shafei [13] showed that for the apparent horizon the first law is
roughly respected for different epochs while the second law of
thermodynamics is respected. Considering the interacting
holographic model of dark energy to investigate the validity of
the GSL of thermodynamics in a non-flat (closed) universe enclosed
by the event horizon, Setare [14] found that generalized second
law is respected for the special range of the deceleration
parameter. The transition from quintessence to phantom dominated
universe was considered and the conditions of the validity of GSL
in transition was studied in [15]. In the reference [16], a
Chaplygin gas dominated was considered and the GSL was
investigated taking into account the existence of the observer's
event horizon in accelerated universes and it was concluded that
for the initial stage of Chaplygin gas dominated expansion, the
GSL of gravitational thermodynamics is fulfilled.\\

In the present work, we study the validity of GSL of
thermodynamics in the \emph{intermediate}[18, 19] and
\emph{logamediate} [19] expansions of the universe bounded by the
Hubble, apparent, particle and event horizons. According to the
GSL, for our system, the sum of the entropy of matter enclosed by
the horizon and the entropy of the horizon must not be a
decreasing function of time. We have investigated the GSL using as
well as without using the first law of thermodynamics. While
considering the GSL we have taken into account the Hubble horizon,
apparent horizon, particle horizon and event horizon.

\section{\normalsize\bf{Generalized Second Law of Thermodynamics}}

We consider the Friedmann-Robertson-Walker (FRW) universe with
line element

\begin{equation}
ds^{2}=-dt^{2}+a^{2}(t)\left[\frac{dr^{2}}{1-kr^{2}}+r^{2}(d\theta^{2}+sin^{2}\theta
d\phi^{2})\right]
\end{equation}

where $a(t)$ is the scale factor and $k$ is the curvature of the
space and $ k = 0, 1$ and $-1$ for flat, closed and open universes
respectively. The Einstein field equations are given by

\begin{equation}
H^{2}+\frac{k}{a^{2}}=\frac{8\pi G}{3}\rho
\end{equation}
and
\begin{equation}
\dot{H}-\frac{k}{a^{2}}=-4\pi G(\rho+p)
\end{equation}

where $\rho$ and $p$ are energy density and isotropic pressure
respectively and $H=\frac{\dot{a}}{a}$ is the Hubble parameter.
The energy conservation equation is given by

\begin{equation}
\dot{\rho}+3H(\rho+p)=0
\end{equation}

We denote the radius of cosmological horizon by $R_{X}$. For
Hubble, apparent, particle and event horizon we replace $X$ by
$H$, $A$, $P$ and $E$ respectively. The corresponding radii are
given by

\begin{equation}
R_{H}=\frac{1}{H}~;~~~R_{A}=\frac{1}{\sqrt{H^{2}+\frac{k}{a^{2}}}}~;~~~
R_{P}=a\int_{0}^{t}\frac{dt}{a}~;~~~R_{E}=a\int_{t}^{\infty}\frac{dt}{a}
\end{equation}

It can be easily obtained that

\begin{equation}
\dot{R}_{H}=-\frac{\dot{H}}{H^{2}}~;~~\dot{R}_{A}=-HR_{A}^{3}\left(\dot{H}-\frac{k}{a^{2}}\right)~;
~~\dot{R}_{P}=HR_{P}+1~;~~\dot{R}_{E}=HR_{E}-1
\end{equation}

 To study the generalized second law (GSL) of thermodynamics through the
universe we deduce the expression for normal entropy using the
Gibb's equation of thermodynamics

\begin{equation}
T_{X}dS_{IX}=pdV_{X}+dE_{IX}
\end{equation}

where, $S_{IX}$ is the internal entropy within the horizon. Here
the expression for internal energy can be written as $E_{IX}=\rho
V_{X}$ ,  where the volume of the sphere is $V_{X}=\frac{4}{3}\pi
R_{X}^{3}$. Using equation (7) we obtain the rate of change of
internal energy as

\begin{equation}
\dot{S}_{IX}=\frac{4\pi
R_{X}^{2}}{T_{X}}(\rho+p)(\dot{R}_{X}-HR_{X})
\end{equation}

In the following, we shall find out the expressions of the rate of
change of total entropy using first law and without using first
law of thermodynamics.\\

\subsection{\large GSL using first law}

Issues related to the said horizons in the context of
thermodynamical studies are available in the references given in
[17]. From the first law of thermodynamics, we have the relation
(Ref. Cai and Kim [5])

\begin{equation}
T_{X}dS_{X}=4\pi R_{X}^{3}H (\rho+p)dt
\end{equation}

where, $T_{X}$ and $R_{X}$ are the temperature and radius of the
horizons under consideration in the equilibrium thermodynamics.\\

Using (9) we can get the time derivative of the entropy on the
horizon as

\begin{equation}
\dot{S}_{X}=\frac{4\pi R_{X}^{3}H}{T_{X}}(\rho+p)
\end{equation}

Adding equations (8) and (10) we get the time derivative of total
entropy as

\begin{equation}
\dot{S}_{X}+\dot{S}_{IX}=\frac{
R_{X}^{2}}{GT_{X}}\left(\frac{k}{a^{2}}-\dot{H}\right)\dot{R}_{X}
\end{equation}

In order the GSL to be hold, we require
$\dot{S}_{X}+\dot{S}_{IX}\geq0$.\\

\subsection{\large GSL without using first law}

In equation (11) the time derivative of the total entropy is
obtained using the first law of thermodynamics. In this paper, we
shall also investigate the GSL without using the first law of
thermodynamics. The horizon entropy is $S_{X}=\frac{\pi
R_{X}^{2}}{G}$ and the temperature is $T_{X}=\frac{1}{2\pi
R_{X}}$. In this case, the time derivative of the entropy on the
horizon is

\begin{equation}
\dot{S}_{X}=\frac{2\pi R_{X}\dot{R}_{X}}{G}
\end{equation}

Therefore, in this case the time derivative of the total entropy
is

\begin{equation}
\dot{S}_{X}+\dot{S}_{IX}=\frac{2\pi
R_{X}}{G}\left[R_{X}^{2}\left(\frac{k}{a^{2}}-\dot{H}\right)\left(\dot{R}_{X}-HR_{X}\right)+
\dot{R}_{X}\right]
\end{equation}

In the following sections, we shall investigate the nature of the
equations (11) and (13) i.e., validity of GS in two scenarios,
namely \emph{logamediate} and \emph{intermediate} scenarios.

\section{\normalsize\bf{GSL in Logamediate Scenario}}

Barrow and Nunes [19] introduced the scenario of `logamediate'
expansion where the scale factor $a(t)$ is given by

\begin{equation}
a(t)=exp(A(\ln t)^{\alpha})~,~~~A\alpha>0~,~~~~\alpha>1~,~~~~t>1
\end{equation}

Subsequently,

\begin{equation}
H=\frac{A\alpha}{t}(\ln t)^{\alpha-1}
\end{equation}

and from equation (5), we obtain

\begin{equation}
\begin{array}{c}
 R_{H}=\frac{t(\ln
t)^{1-\alpha}}{A\alpha}~;~~R_{A}=\frac{1}{\sqrt{e^{-2A(\ln
t)^{\alpha}}k+\frac{A^{2}\alpha^{2}(\ln
t)^{-2(1-\alpha)}}{t^{2}}}}~;~~ \\\\
  R_{P}=exp(A(\ln
t)^{\alpha})\int_{0}^{t}\frac{dt}{exp(A(\ln
t)^{\alpha})}~;~~R_{E}=exp(A(\ln
t)^{\alpha})\int_{t}^{\infty}\frac{dt}{exp(A(\ln t)^{\alpha})} \\
\end{array}
\end{equation}

Using (11), (13)-(16) we get the time derivatives of the total
entropies to investigate the validity of the GSL in various
horizons using and without using first law.\\

\subsection{\bf GSL in the \emph{logamediate} scenario using first law}

Here we consider the GSL in the \emph{logamediate} scenario using
the first law of thermodynamics. Using (11) and (16) we get the
time derivative of total entropies as follows:\\
\begin{itemize}
    \item For Hubble horizon
    \begin{equation}
    \dot{S}_{H}+\dot{S}_{IH}=\frac{e^{-2A(\ln t)^{\alpha}}(\alpha-1-\ln t)(\ln t)^{-3\alpha}\left(-kt^{2}(\ln t)^{2}
    +A\alpha e^{2A(\ln t)^{\alpha}}(\ln t)^{\alpha}(\alpha-1-\ln t)   \right)}{A^{3}\alpha^{3}GT_{H}}
    \end{equation}
    \item For apparent horizon
    \begin{equation}
     \dot{S}_{A}+\dot{S}_{IA}=\frac{A\alpha e^{A(\ln t)^{\alpha}}(\ln t)^{\alpha}\left(kt^{2}(\ln t)^{2}
     -Ae^{2A(\ln t)^{\alpha}}\alpha(\alpha-1-\ln t)(\ln t)^{\alpha}\right)^{2}}{GT_{A}\left(kt^{2}(\ln t)^{2}
     +A^{2}e^{2A(\ln t)^{\alpha}}\alpha^{2}(\ln
     t)^{2\alpha}\right)^{5/2}}~~~~~~~~~~~~~~~~~~~~~\\
    \end{equation}
    \item For particle horizon
    \begin{equation}
    \dot{S}_{P}+\dot{S}_{IP}=\frac{e^{-2A(\ln t)^{\alpha}} R_{P}^{2}\left(t\ln t+A \alpha (\ln t)^{\alpha}R_{P}\right)
    (kt^{2}(\ln t)^{2}-Ae^{2A(\ln t)^{\alpha}}\alpha(\alpha-1-\ln t)(\ln t)^{\alpha})}{t^{3} (\ln t)^{3}GT_{P}}\\
      \end{equation}
    \item For event horizon
    \begin{equation}
    \dot{S}_{E}+\dot{S}_{IE}=\frac{e^{-2A(\ln t)^{\alpha}} R_{E}^{2}\left(-t\ln t+A \alpha (\ln t)^{\alpha}R_{E}\right)
    (kt^{2}(\ln t)^{2}-Ae^{2A(\ln t)^{\alpha}}\alpha(\alpha-1-\ln t)(\ln t)^{\alpha})}{t^{3} (\ln t)^{3}GT_{E}}\\
      \end{equation}
\end{itemize}

In figures 1-4 we have investigated the GSL using the first law of
thermodynamics. Here also the time derivative of the horizon
entropy is calculated using the first law of thermodynamics. In
figures 1 and 2 we have plotted the time derivatives of the total
entropies against the cosmic time $t$ for the Hubble and apparent
horizons in the \emph{logamediate} scenario. We have computed the
total entropy using the first law of thermodynamics. We see that
the rate of change of total entropies are decreasing with time.
We find that $\dot{S}_{X}+\dot{S}_{IX}>0$ throughout the evolution
of the universe. Thus, the GSL is valid in both of the Hubble and
apparent horizons.\\

In figure 3, we have plotted the time derivative of the total
entropy assuming particle horizon as the enveloping horizon of the
universe. We see that the rate of change of total entropy is
increasing with time. In this figure the time derivative stays at
the positive level. This indicates that the GSL is valid on the
particle horizon. Also, in figure 4, where the time derivative of
the total entropy is plotted for the universe enveloped by the
event horizon, we see that the rate of change of total entropy is
decreasing with time. We find that $\dot{S}_{E}+\dot{S}_{IE}>0$
throughout the evolution of the universe. This indicates that when
we consider the cosmological event horizon as the enveloping
horizon of the universe, the GSL is also valid. So we conclude
that the GSL is always valid for Hubble, apparent, particle and
event horizons in the \emph{logamediate} scenario of the universe
when we have calculated the horizon entropies using first law
of thermodynamics. \\

\begin{figure}
\includegraphics[height=2in]{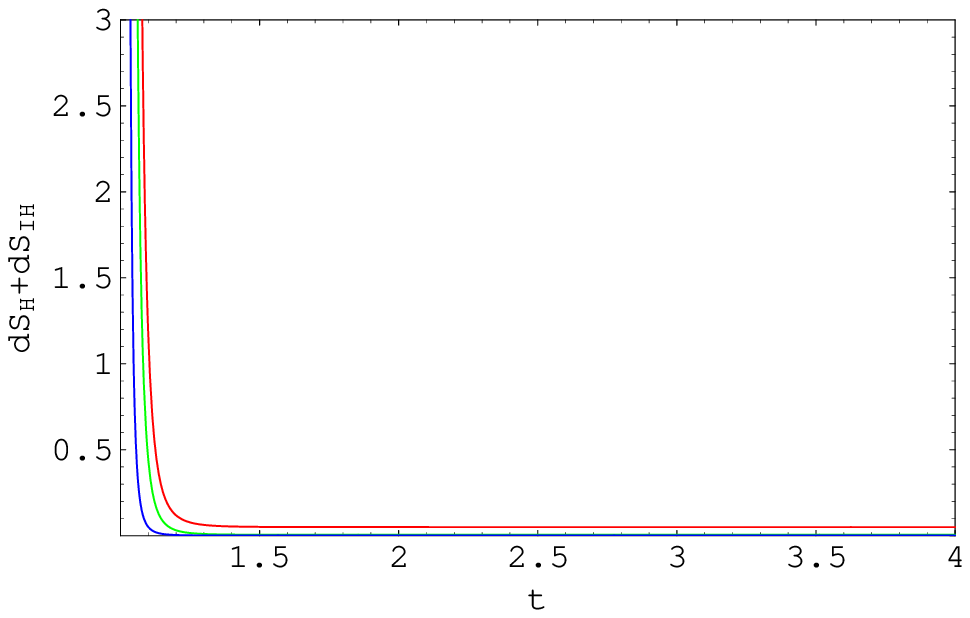}~~~~
\includegraphics[height=2in]{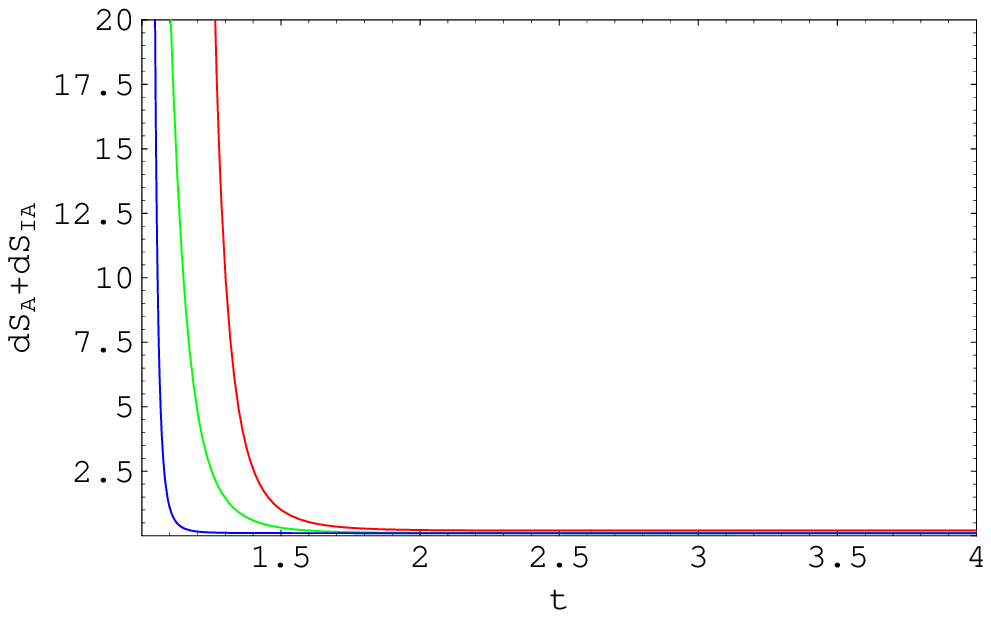}\\
\vspace{1mm} ~~~~~~~Fig.1~~~~~~~~~~~~~~~~~~~~~~~~~~~~~~~~~~~~~~~~~~~~~~~~~~~~~~~~~~~~~~~~~~~~~~~~~Fig.2~~~\\

\vspace{6mm}

\includegraphics[height=2in]{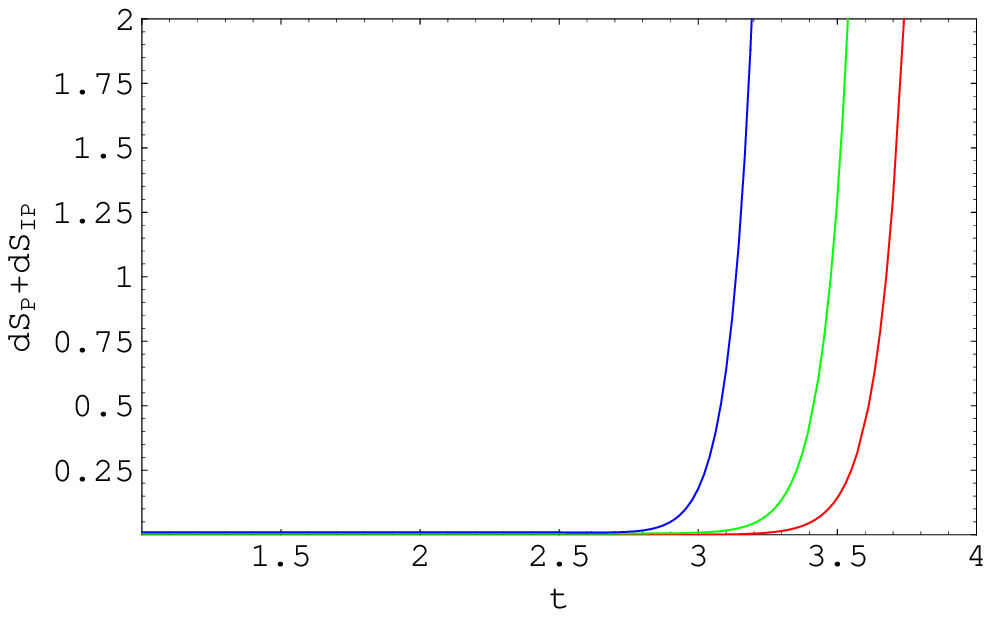}~~~~
\includegraphics[height=2in]{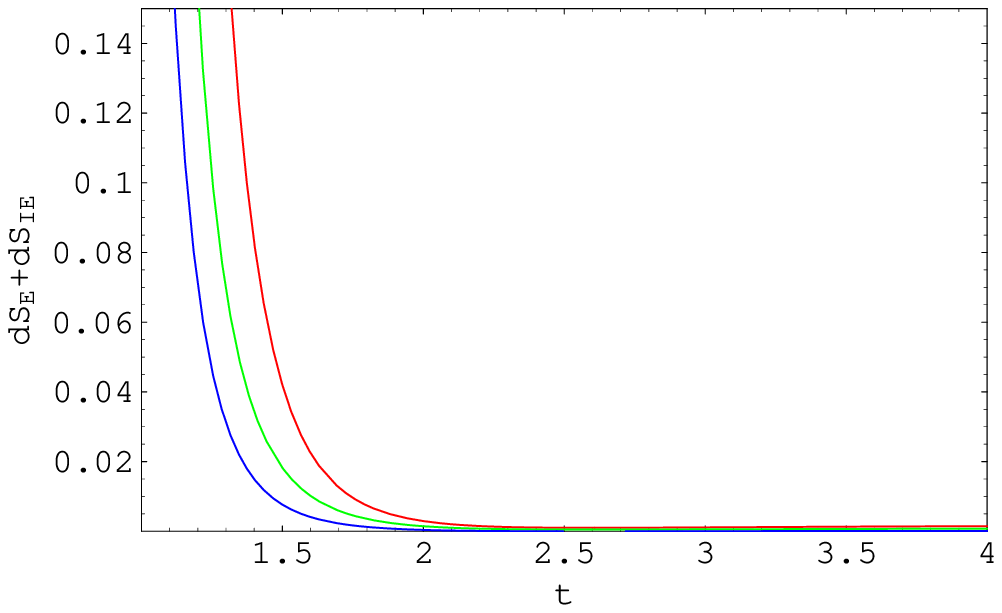}\\
\vspace{1mm} ~~~~~~~Fig.3~~~~~~~~~~~~~~~~~~~~~~~~~~~~~~~~~~~~~~~~~~~~~~~~~~~~~~~~~~~~~~~~~~~~~~~~~Fig.4~~~\\

\vspace{6mm} Figs. 1, 2, 3 and 4 show the time derivatives of the
total entropy for Hubble horizon $R_{H}$, apparent horizon
$R_{A}$, particle horizon $R_{P}$ and event horizon $R_{E}$
respectively \textbf{using first law of thermodynamics} in the
\emph{logamediate} scenario. The red, green and blue lines
represent the $dS_{X}+dS_{IX}$ for $k=-1,~1$ and $0$
respectively. We have chosen $A=5,~~\alpha=2$.\\

 \vspace{6mm}

 \end{figure}

\begin{figure}
 \includegraphics[height=2in]{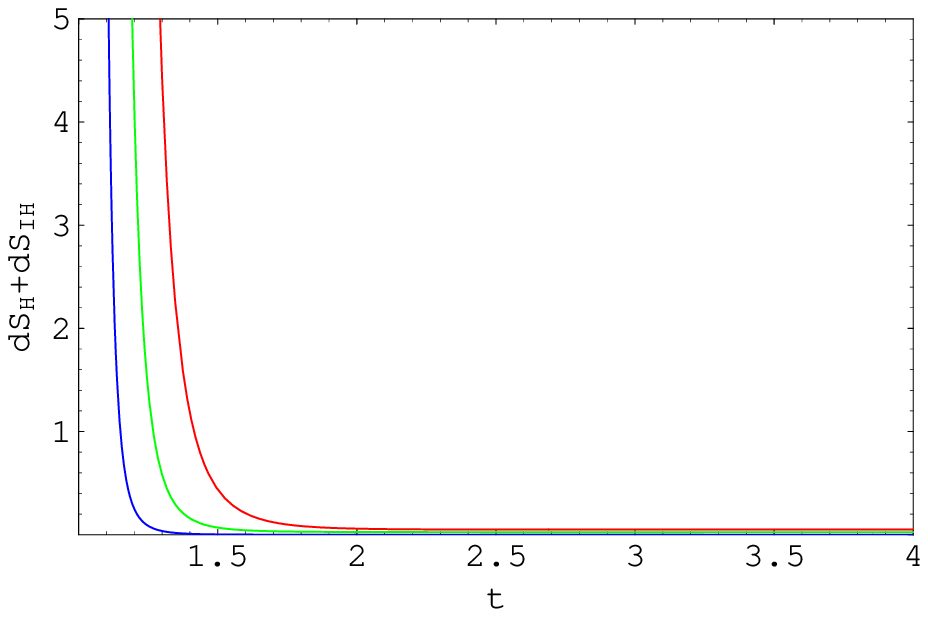}~~~~
\includegraphics[height=2in]{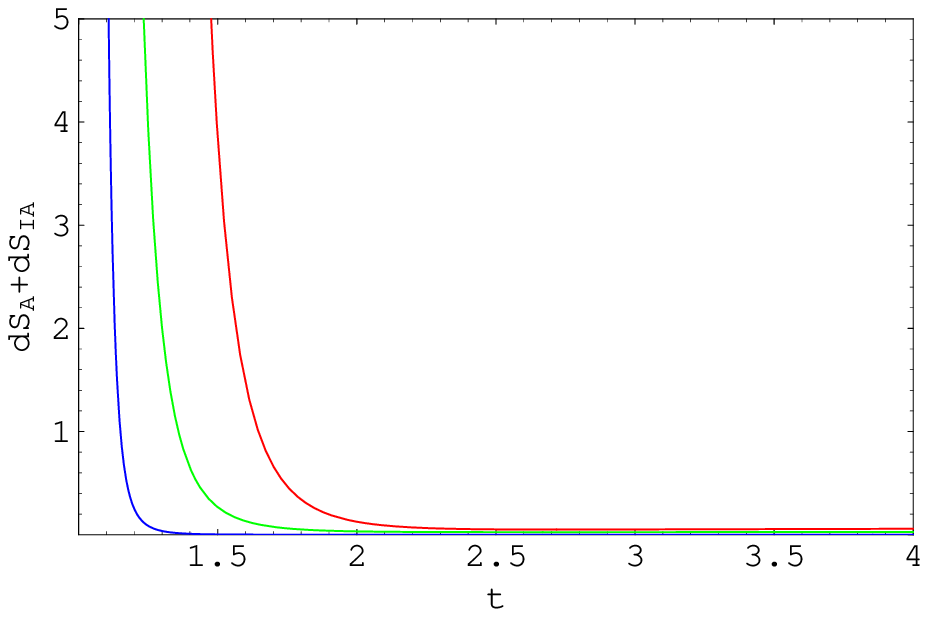}\\
\vspace{1mm} ~~~~~~~Fig.5~~~~~~~~~~~~~~~~~~~~~~~~~~~~~~~~~~~~~~~~~~~~~~~~~~~~~~~~~~~~~~~~~~~~~~~~~Fig.6~~~\\

\vspace{6mm}
\includegraphics[height=2in]{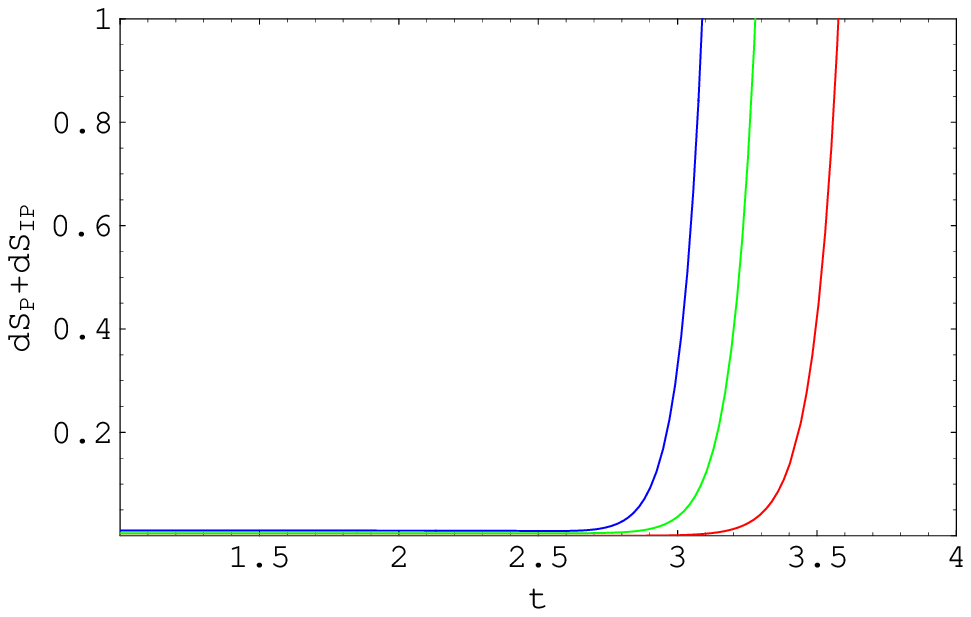}~~~~
\includegraphics[height=2in]{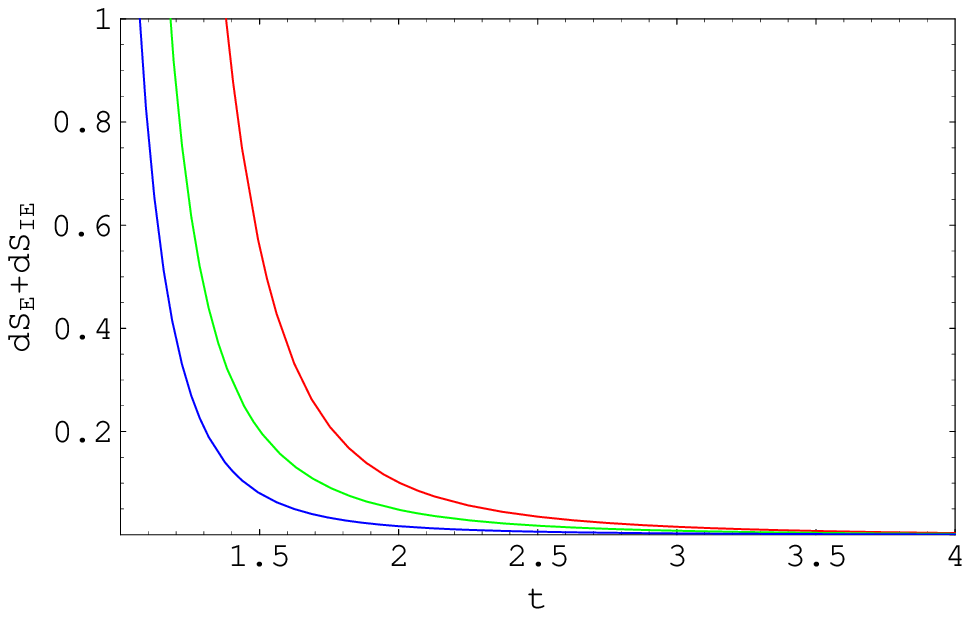}\\
\vspace{1mm} ~~~~~~~Fig.7~~~~~~~~~~~~~~~~~~~~~~~~~~~~~~~~~~~~~~~~~~~~~~~~~~~~~~~~~~~~~~~~~~~~~~~~~Fig.8~~~\\

\vspace{6mm} Figs. 5, 6, 7 and 8 show the time derivatives of the
total entropy for Hubble horizon $R_{H}$, apparent horizon
$R_{A}$, particle horizon $R_{P}$ and event horizon $R_{E}$
respectively \textbf{without using first law of thermodynamics} in
the \emph{logamediate} scenario. The red, green and blue lines
represent the $dS_{X}+dS_{IX}$ for $k=-1,~1$ and $0$
respectively. We have chosen $A=5,~~\alpha=2$.\\

 \vspace{3mm}

 \end{figure}

\subsection{\bf GSL in the \emph{logamediate} scenario without using
first law}

Without using the first law, the time derivative of the total
entropies in the \emph{logamediate} scenario come out as

\begin{itemize}
    \item For Hubble horizon
    \begin{equation}
    \dot{S}_{H}+\dot{S}_{IH}=\frac{2e^{-2A(\ln t)^{\alpha}}\pi t (\ln t)^{1-4\alpha}\left(kt^{2}(\ln t)^{2}
    (1-\alpha+\ln t)+A\alpha (\ln t)^{\alpha}(-kt^{2}(\ln t)^{2}+e^{2A(\ln t)^{\alpha}}(1-\alpha+\ln t)^{2})\right)}{A^{4}\alpha^{4}G}
    \end{equation}
    \item For apparent horizon
    \begin{equation}
     \dot{S}_{A}+\dot{S}_{IA}=\frac{2Ae^{2A(\ln t)^{\alpha}}\pi t\alpha (\ln t)^{1+\alpha}\left(kt^{2}(\ln t)^{2}
     -Ae^{2A(\ln t)^{\alpha}}\alpha(\alpha-1-\ln t)(\ln t)^{\alpha}\right)^{2}}{G\left(kt^{2}(\ln t)^{2}+A^{2}\alpha^{2}e^{2A(\ln t)^{\alpha}}(\ln
     t)^{2\alpha}\right)^{3}}~~~~~~~~~~~~~~~~~~~~\\
    \end{equation}
    \item For particle horizon
    \begin{equation}
    \dot{S}_{P}+\dot{S}_{IP}=\frac{2\pi R_{P}^{3}}{G}\left(ke^{-2A(\ln t)^{\alpha}}-\frac{A\alpha(\alpha-1)(\ln t)^{-2+\alpha}}{t^{2}}
    +\frac{A\alpha (\ln t)^{\alpha-1}}{t^{2}}\right)+\frac{2\pi R_{P}}{G}\left(\frac{A\alpha R_{P}(\ln t)^{\alpha-1}}{t}+1\right)
    \end{equation}
    \item For event horizon
    \begin{equation}
      \dot{S}_{E}+\dot{S}_{IE}=-\frac{2\pi R_{E}^{3}}{G}\left(ke^{-2A(\ln t)^{\alpha}}-\frac{A\alpha(\alpha-1)(\ln t)^{-2+\alpha}}{t^{2}}
      +\frac{A\alpha (\ln t)^{\alpha-1}}{t^{2}}\right)+\frac{2\pi R_{E}}{G}\left(\frac{A\alpha R_{E}(\ln t)^{\alpha-1}}{t}-1\right)
    \end{equation}
    \end{itemize}

In figures 5 - 8 we have plotted $\dot{S}_{X}+\dot{S}_{IX}$ based
on equations (21) - (24) against cosmic time $t$ without using the
first law of thermodynamics in the \emph{logamediate} scenario.
Here also we find similar results to those obtained using the
first law of thermodynamics. From figures 5-8 we find that the GSL
is valid in the \emph{logamediate} scenario without using the
first law for the universe enveloped by the Hubble, apparent,
particle and event horizons.\\

\section{\normalsize\bf{GSL in Intermediate Scenario}}

Barrow and Liddle [18] proposed a model of `intermediate'
expansion, where the scale factor is given by

\begin{equation}
a(t)=exp (B t^{\beta})~,~~~B>0,~~0<\beta<1
\end{equation}

This model bears many qualitative similarities to power-law
inflation: like power-law inflation, there is no natural end to
inflation and a mechanism must be introduced in order to bring
inflation to an end. Also, as with power-law inflation,
intermediate inflation offers the possibility of density
perturbation and gravitational wave spectra which differ
significantly from the usual inflationary prediction of a nearly
flat spectrum with negligible gravitational waves [18]. Using the
above form of scale factor, we get

\begin{equation}
H=B\beta t^{\beta-1}
\end{equation}

Subsequently from equation (5), we obtain,

\begin{equation}
\begin{array}{c}
R_{H}=\frac{t^{1-\beta}}{B\beta}~;~~R_{A}=\frac{1}{\sqrt{ke^{-2Bt^{\beta}}+B^{2}\beta^{2}t^{2(\beta-1)}}}~;\\\\
R_{P}=\frac{B^{-\frac{1}{\beta}}e^{Bt^{\beta}}\left(\Gamma\left[\frac{1}{\beta}\right]-\Gamma\left[\frac{1}{\beta},Bt^{\beta}\right]\right)}{\beta}~;
~~R_{E}=\frac{B^{-\frac{1}{\beta}}e^{Bt^{\beta}}\Gamma\left[\frac{1}{\beta},Bt^{\beta}\right]}{\beta} \\
\end{array}
\end{equation}

Using the above expressions in equations (11) and (13) we can get
the time derivatives of the total entropies using as well as
without using the first law of thermodynamics.\\

\begin{figure}
 \includegraphics[height=2in]{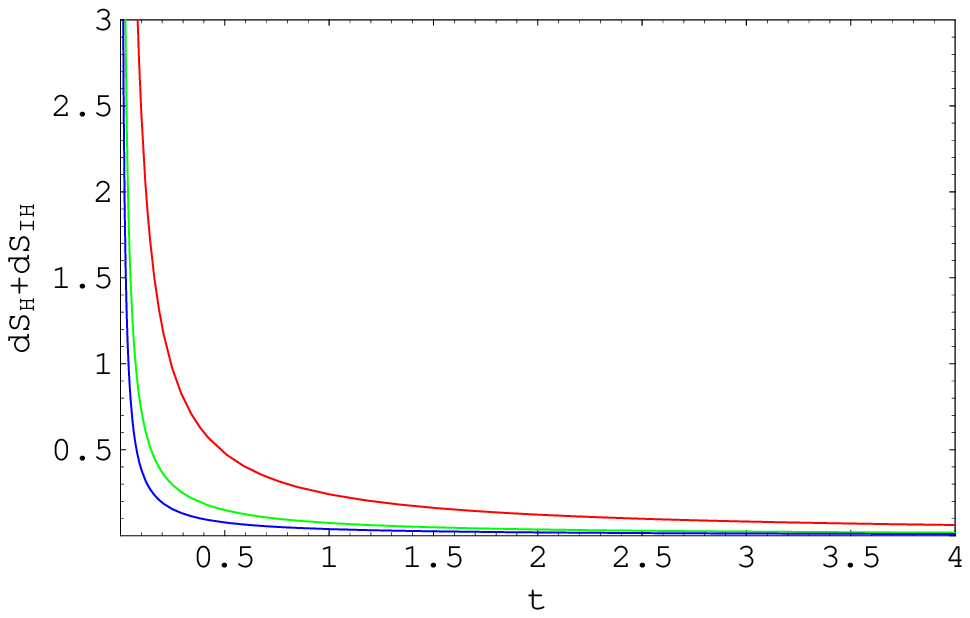}~~~~
\includegraphics[height=2in]{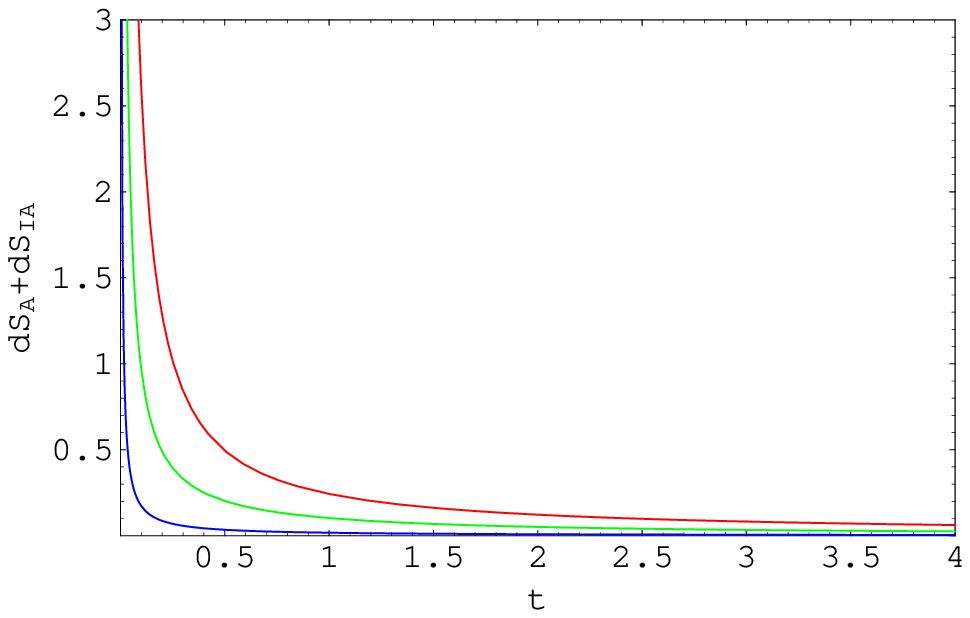}\\
\vspace{1mm} ~~~~~~~Fig.9~~~~~~~~~~~~~~~~~~~~~~~~~~~~~~~~~~~~~~~~~~~~~~~~~~~~~~~~~~~~~~~~~~~~~~~~~Fig.10~~~\\

\vspace{6mm}
\includegraphics[height=2in]{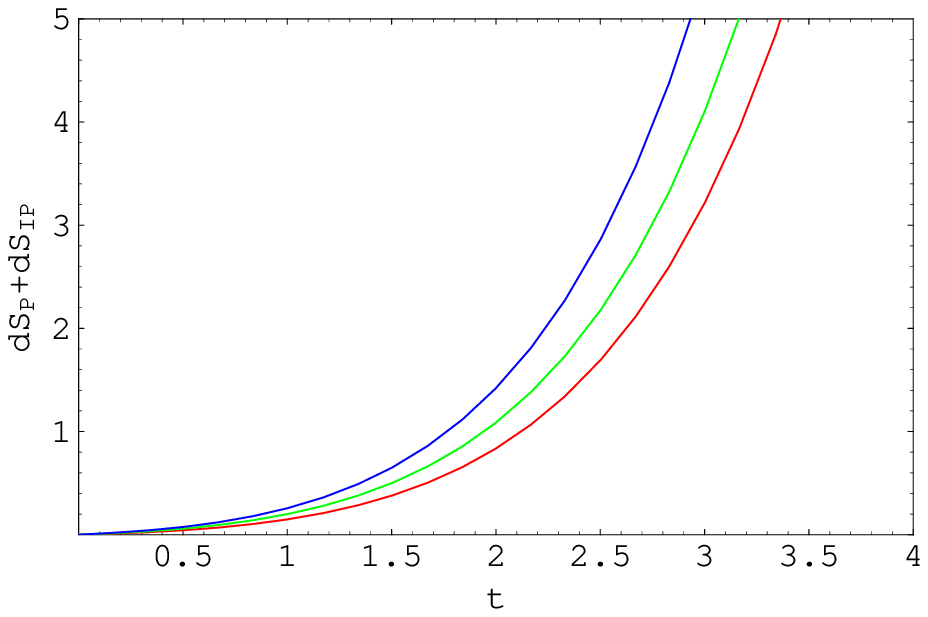}~~~~
\includegraphics[height=2in]{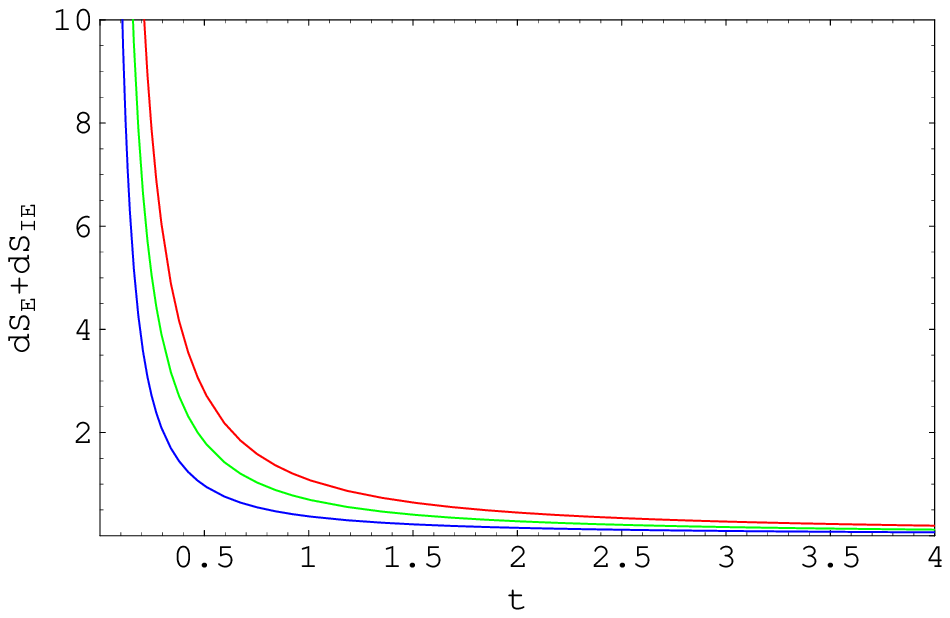}\\
\vspace{1mm} ~~~~~~~Fig.11~~~~~~~~~~~~~~~~~~~~~~~~~~~~~~~~~~~~~~~~~~~~~~~~~~~~~~~~~~~~~~~~~~~~~~~~~Fig.12~~~\\

\vspace{6mm} Figs. 9, 10, 11 and 12 show the time derivatives of
the total entropy for Hubble horizon $R_{H}$, apparent horizon
$R_{A}$, particle horizon $R_{P}$ and event horizon $R_{E}$
respectively \textbf{using first law of thermodynamics} in the
\emph{intermediate} scenario. The red, green and blue lines
represent the $dS_{X}+dS_{IX}$ for $k=-1,~1$ and $0$
respectively. We have chosen $A=5,~~\alpha=2$.\\

 \vspace{6mm}

 \end{figure}

\subsection{\bf GSL in the \emph{intermediate }scenario using first
law}

In this subsection we consider the GSL in the \emph{intermediate}
scenario. Using the first law the time derivatives of the total
entropies are \\

\begin{itemize}
    \item For Hubble horizon
    \begin{equation}
    \dot{S}_{H}+\dot{S}_{IH}=\frac{e^{-2Bt^{\beta}} t^{-3\beta}(\beta-1)\left(-kt^{2}
    +Be^{2Bt^{\beta}}t^{\beta}(\beta-1)\beta\right)}{B^{3}\beta^{3}GT_{H}}
    \end{equation}
    \item For apparent horizon
    \begin{equation}
    \dot{S}_{A}+\dot{S}_{IA}=\frac{B\beta e^{Bt^{\beta}}t^{\beta}\left(kt^{2}-Be^{2Bt^{\beta}}t^{\beta}
    (\beta-1)\beta\right)^{2}}{GT_{A}\left(kt^{2}+B^{2}e^{2Bt^{\beta}}t^{2\beta}\beta^{2}\right)^{5/2}}
    \end{equation}
    \item For particle horizon
      \begin{equation}
      \dot{S}_{P}+\dot{S}_{IP}=\frac{B^{-\frac{3}{\beta}}\left(\Gamma\left[\frac{1}{\beta}\right]
      -\Gamma\left[\frac{1}{\beta},Bt^{\beta}\right]\right)^{2}
      \left(kt^{2}-Be^{2Bt^{\beta}}t^{\beta}(\beta-1)\beta\right)\left(B^{\frac{1}{\beta}}t+Be^{Bt^{\beta}}
      t^{\beta}\left(\Gamma\left[\frac{1}{\beta}\right]
      -\Gamma\left[\frac{1}{\beta},Bt^{\beta}\right]\right)\right)}{t^{3}\beta^{2}GT_{P}}\\
      \end{equation}
      \item   For event horizon
  \begin{equation}
   \dot{S}_{E}+\dot{S}_{IE}=\frac{B^{-\frac{3}{\beta}}\left(\Gamma\left[\frac{1}{\beta},Bt^{\beta}\right]\right)^{2}
      \left(kt^{2}-Be^{2Bt^{\beta}}t^{\beta}(\beta-1)\beta\right)\left(-B^{\frac{1}{\beta}}t+Be^{Bt^{\beta}}
      t^{\beta}\Gamma\left[\frac{1}{\beta},Bt^{\beta}\right]\right)}{t^{3}\beta^{2}GT_{E}}
    \end{equation}
\end{itemize}

In figures 9 - 12 we have investigated the GSL for the
\emph{intermediate} scenario using the first law of thermodynamics
using equations (28) - (31). Like the \emph{logamediate} scenario,
the GSL in this situation is valid for the universes enveloped by
Hubble, apparent and event horizon. Also, in all these cases, the
time derivatives of the total entropy are falling with the passage
of cosmic time. However, in the case of the universe enveloped by
the apparent horizon, the time derivative of the total entropy is
positive and increasing throughout the evolution of the universe.
In this case also the GSL is satisfied. So we may conclude that
the GSL is valid for all horizons in intermediate scenario of the
universe using first law.\\

\subsection{\bf GSL in the \emph{intermediate} scenario without using
first law}

The time derivatives of the total entropies are also calculated
without using the first law of thermodynamics in the
\emph{intermediate} scenario as follows:\\

 \begin{figure}
 \includegraphics[height=2in]{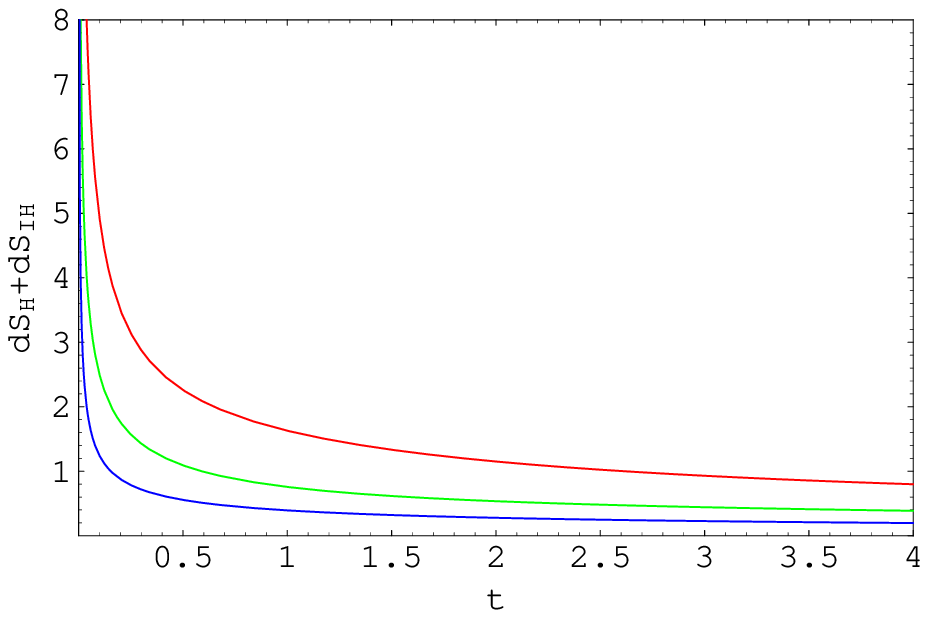}~~~~
\includegraphics[height=2in]{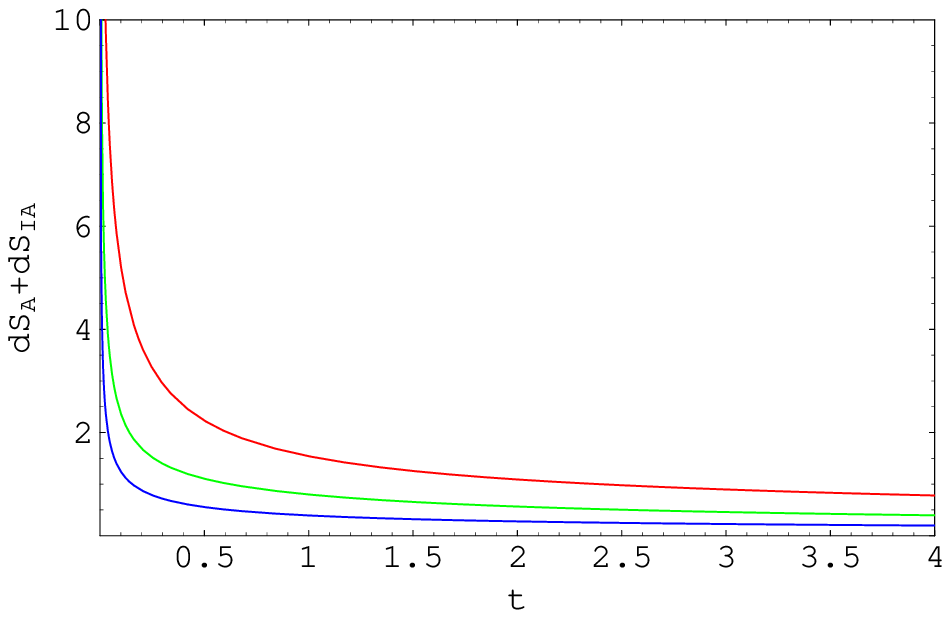}\\
\vspace{1mm} ~~~~~~~Fig.13~~~~~~~~~~~~~~~~~~~~~~~~~~~~~~~~~~~~~~~~~~~~~~~~~~~~~~~~~~~~~~~~~~~~~~~~~Fig.14~~~\\

\vspace{6mm}
\includegraphics[height=2in]{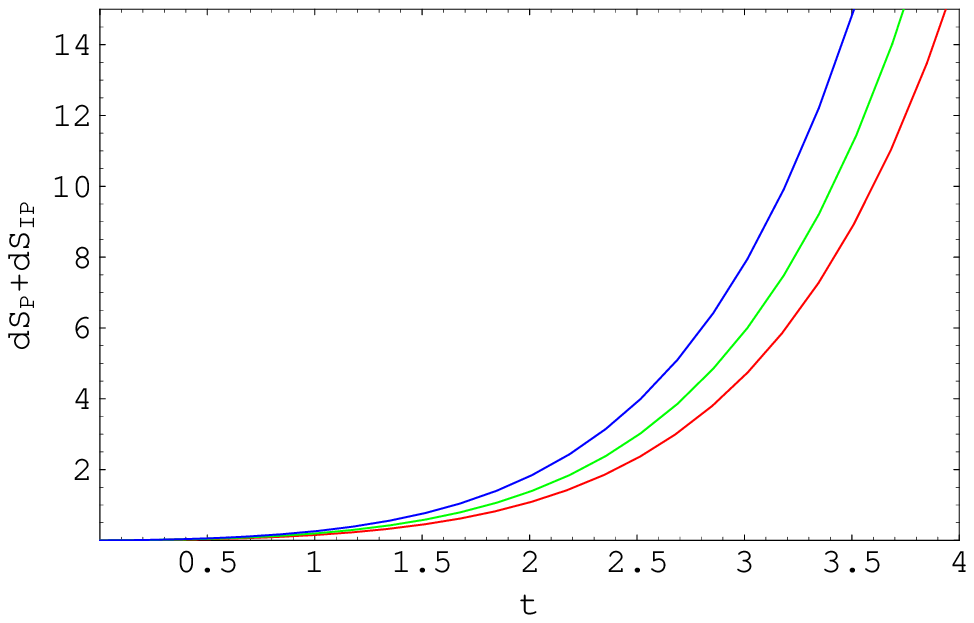}~~~~
\includegraphics[height=2in]{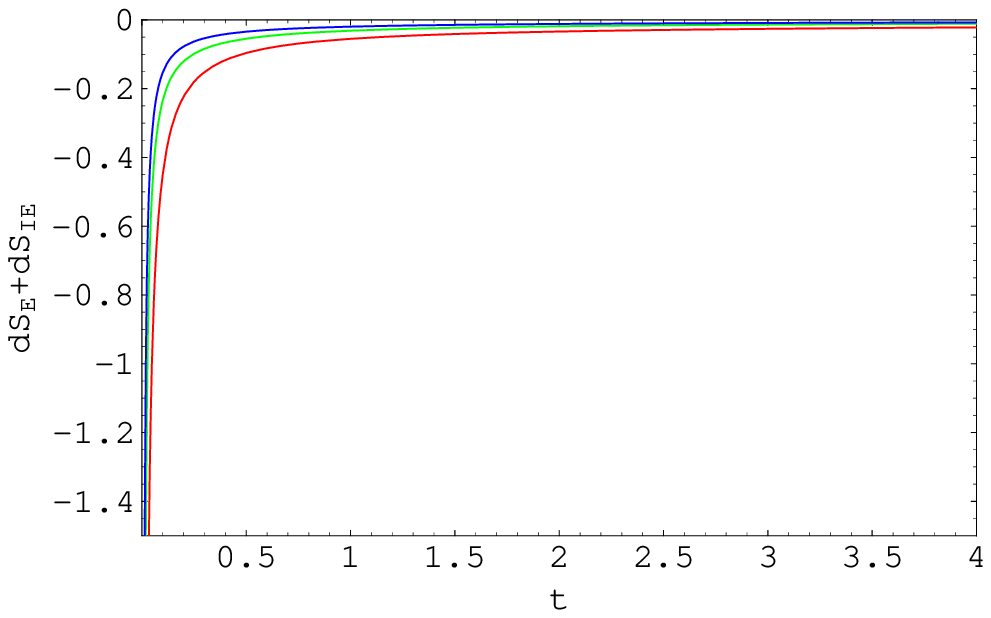}\\
\vspace{1mm} ~~~~~~~Fig.15~~~~~~~~~~~~~~~~~~~~~~~~~~~~~~~~~~~~~~~~~~~~~~~~~~~~~~~~~~~~~~~~~~~~~~~~~Fig.16~~~\\

\vspace{6mm} Figs. 13, 14, 15 and 16 show the time derivatives of
the total entropy for Hubble horizon $R_{H}$, apparent horizon
$R_{A}$, particle horizon $R_{P}$ and event horizon $R_{E}$
respectively \textbf{without using first law of thermodynamics} in
the \emph{intermediate} scenario. The red, green and blue lines
represent the $dS_{X}+dS_{IX}$ for $k=-1,~1$ and $0$
respectively. We have chosen $A=5,~~\alpha=2$.\\

\vspace{6mm}

\end{figure}

\begin{itemize}
    \item For Hubble horizon
    \begin{equation}
    \dot{S}_{H}+\dot{S}_{IH}=\frac{2\pi e^{-2Bt^{\beta}}
    t^{1-4\beta}\left(Be^{2Bt^{\beta}}t^{\beta}\beta(\beta-1)^{2}-kt^{2}(\beta-1(1+Bt^{\beta}))
    \right)}{B^{4}\beta^{4}G}
    \end{equation}
    \item For apparent horizon
    \begin{equation}
    \dot{S}_{A}+\dot{S}_{IA}=\frac{2\pi\beta Be^{2Bt^{\beta}}t^{1+\beta} \left(kt^{2}-Be^{2Bt^{\beta}}t^{\beta}
    (\beta-1)\beta\right)^{2}}{G\left(kt^{2}+B^{2}\beta^{2}e^{2Bt^{\beta}}t^{2\beta}\right)^{3}}
    \end{equation}
    \item For particle horizon
       \begin{eqnarray*}
      \dot{S}_{P}+\dot{S}_{IP}=\frac{2\pi B^{-\frac{1}{\beta}}e^{Bt^{\beta}}\left(\Gamma\left[\frac{1}{\beta}\right]
      -\Gamma\left[\frac{1}{\beta},Bt^{\beta}\right]\right)}{\beta^{3}G}\times
       \left[\beta^{2}+B^{1-\frac{1}{\beta}}\beta^{2}e^{Bt^{\beta}}t^{\beta-1}
       \left(\Gamma\left[\frac{1}{\beta}\right]-\Gamma\left[\frac{1}{\beta},Bt^{\beta}\right]\right)\right.
    \end{eqnarray*}
    \begin{equation}
~~~~~~~~~~~~~~~~~\left.+B^{-\frac{2}{\beta}}
\left(k-Be^{2Bt^{\beta}}t^{\beta-2}(\beta-1)\beta\left(\Gamma
       \left[\frac{1}{\beta}\right]-\Gamma\left[\frac{1}{\beta},Bt^{\beta}\right]\right)^{2}\right)\right]
    \end{equation}
    \item For event horizon
    \begin{eqnarray*}
    \dot{S}_{E}+\dot{S}_{IE}=
    \frac{2\pi
    B^{-\frac{3}{\beta}}e^{Bt^{\beta}}\Gamma\left[\frac{1}{\beta},Bt^{\beta}\right]}{t^{2}\beta^{3}G}\times
    \left[\beta Be^{Bt^{\beta}}t^{\beta}\left(\beta B^{\frac{1}{\beta}}t+(\beta-1)e^{Bt^{\beta}}
    \Gamma\left[\frac{1}{\beta},Bt^{\beta}\right]\right) \right.
    \end{eqnarray*}
 \begin{equation}
~~~~~~~~~~~~~~~~~~~~ \left.
-t^{2}\left(\beta^{2}B^{\frac{2}{\beta}}+k
    \left(\Gamma\left[\frac{1}{\beta},Bt^{\beta}\right]\right)^{2}  \right) \right]
    \end{equation}
\end{itemize}

Time derivatives of the total entropies are calculated based on
the equations (32) - (35) and are plotted against the cosmic time
$t$. Figures 13 - 15 show the staying of the time derivative of
the total entropy at positive level throughout the evolution of
the universe. This indicates the validity of the GSL in the
universes enveloped by Hubble, apparent and particle horizons.
However, it fails to stay at positive level in the case of the
universe enveloped by the event horizon. This indicates the
breaking down of the GSL without using the first law of
thermodynamics in the \emph{intermediate} scenario.\\

\section{Discussions}

In the present work, our endeavor was to investigate the validity
of the generalized second law of thermodynamics in the
\emph{logamediate} and \emph{intermediate} scenarios of the
universe bounded by the Hubble, apparent, particle and event
horizons. We have investigated the generalized second law using
two different approaches namely, (i) using the first law of
thermodynamics and (ii) without using the first law of thermodynamics.\\

As previously stated, the basic aim is to investigate whether
$dS_{X}+dS_{IX}$ i.e. the time derivative of the sum of the
entropy of the universe bounded by the horizons and the normal
entropy remains at non-negative level. To do the same, we have
considered the universe enveloped by Hubble, apparent, particle
and event horizons. The logamediate as well as intermediate
scenarios have been investigated for the nature of the time
derivative of the total entropy for four of the said horizons.
With suitable choice of the model parameters, we have seen from
figures 1 - 4 that if we use the first law to derive the time
derivative of the total entropy, the generalized second law is
valid for all of the four horizons irrespective of the curvature
of the universe. Similar situation happens (see figures 5 - 8)
when we ignore the first law to get the time derivative of the
total entropy. While considering the intermediate scenario we find
from figures 9 - 12 that the generalized second law based on first
law is always valid for all types of enveloping horizons and the
curvature of the universe. However, if we ignore the first law, we
find that in intermediate scenario the generalized second law
breaks down (see figure 16) for open, flat and closed universe
enveloped by the event horizon. The generalized second law without
the first law in this scenario is valid if the enveloping horizons
are Hubble, apparent and particle horizons (see figures 12 - 15).
The validity of the generalized second law occurs irrespective of
the use of the first law of thermodynamics in calculating the time
derivative of the total entropy.\\\\

{\bf References:}\\
\\
$[1]$ S. J. Perlmutter et al., {\it Astrophys. J.} {\bf 517} 565
(1999); A. G. Reiss et al., {\it Astron. J.} {\bf 116} 1009
(1998); J. L. Tonry et al., {\it Astrophys. J.} {\bf 594} 1
(2003); B. J. Barris, {\it Astrophys. J.} {\bf 602} 571 (2004); A.
G. Reiss et al., {\it Astrophys. J.} {\bf 607} 665 (2004).
\\\\
$[2]$ T. Padmanabhan, {\it Current Science} {\bf 88} 1057
(2005).\\\\
$[3]$ M. Tegmark et al., {\it Phys. Rev. D} {\bf 74} 123507
(2006); S. Perlmutter et al., {\it Astrophys. J.} {\bf 517} 565
(1999).
\\\\
$[4]$ V. Sahni and A. A. Starobinsky, {\it Int. J. Mod. Phys. D}
{\bf 9} 373 (2000); T. Padmanabhan, {\it Phys. Rep.} {\bf 380} 235
(2003);  T. Padmanabhan, {\it Current Science} {\bf 88} 1057
(2005); T. Padmanabhan, {\it Advanced Science Letters} {\bf 2} 174
(2009); E. J. Copeland, M. Sami, and S. Tsujikawa, {\it Int. J.
Mod. Phys. D} {\bf 15} 1753 (2006).
\\\\
$[5]$ H. M. Sadjadi, {\it Phys. Rev. D} {\bf 76} 104024 (2007); R.
Bousso, {\it Phys. Rev. D} {\bf 71} 064024 (2005); S. Nojiri and
S. D. Odintsov, {\it Phys. Rev. D} {\bf 70} 103522 (2004); Y. S.
Piao, {\it Phys. Rev. D} {\bf 74} 047301 (2006); T. Jacobson, {\it
Phys. Rev. Lett.} {\bf 75} 1260 (1995); R. G. Cai and S. P. Kim,
{\it JHEP} {\bf 0502} 050 (2005); Y. Gong, B. Wang and A. Wang ,
{\it Phys. Rev. D} {\bf 75} 123516 (2007); T. M. Davis, P. C. W.
Davies and C. H. Lineweaver, {\it Classical Quantum Gravity} {\bf
20} 2753 (2003).\\\\
$[6]$ T. Jacobson, {\it Phys. Rev. Lett.} {\bf 75} 1260
(1995).\\\\
$[7]$ E. Verlinde, hep-th/0008140.\\\\
$[8]$ R. G. Cai and S. P. Kim, {\it JHEP} {\bf 02} 050 (2005); Y.
Gong and A. Wang, {\it Phys. Rev. Lett.} {\bf 99} 211301 (2007);
R-G.  Cai and L-M. Cao, {\it Phys. Rev. D} {\bf 75} 064008 (2007);
X-H. Ge,  {\it Phys. Lett. B} {\bf 651} 49 (2007); 6.  M. Akbar
and R-G.  Cai, {\it Phys. Lett. B} {\bf 635} 7 (2006).
\\\\
$[9]$  R. Guedens and T. Jacobson, {\it Phys. Rev. Lett.} {\bf 96}
121301 (2006).\\\\
$[10]$ Y. Gong and A. Wang, {\it Phys. Rev. Lett.} {\bf 99} 211301 (2007).\\\\
$[11]$ B. Wang, Y. G. Gong and E. Abdalla, {\it Phys. Rev. D} {\bf
74} 083520
(2006).\\\\
$[12]$ L. N. Granda and A. Oliveros, {\it Phys. Lett. B} {\bf 669}
275 (2008); Y. Gong, B. Wang and A. Wang, {\it JCAP} {\bf 01} 024
(2007); T. Padmanabhan, {\it Class. Quantum Grav.} {\bf 19} 5387
(2002); R. -G. Cai and N. Ohta, {\it Phys. Rev. D} {\bf 81}
084061; R. G. Cai and L. -M. Cao, {\it Nucl. Phys. B} {\bf 785}
135 (2007); M. Akbar and R. -G. Cai, {\it Phys. Lett. B} {\bf 635}
7 (2006); R. -G. Cai, L. -M. Cao, Y. -P. Hu and S. P. Kim, {\it
Phys. Rev. D} {\bf 78} 124012 (2008).\\\\
$[13]$ M. R. Setare and S. Shafei, {\it JCAP} {\bf 09} 011
(2006).\\\\
$[14]$ M. R. Setare, {\it JCAP} {\bf 01} 023 (2007).\\\\
$[15]$ M. R. Setare, {\it Phys. Lett. B} {\bf 641} 130 (2006).\\\\
$[16]$ G. Izquierdo and D. Pavon, {\it Phys. Lett. B} {\bf 633}
420 (2006).\\\\
$[17]$ M. Li, {\it Phys. Lett. B} {\bf 603} 1 (2004); S. Nojiri
and S. D. Odintsov, {\it Gen. Rel. Grav.} {\bf 38} 1285 (2006); R.
Easther and D. A. Lowe, {\it Phys. Rev. Lett.} {\bf 82} 4967
(1999).
\\\\
$[18]$ J. D. Barrow and A. R. Liddle, {\it Phys. Rev. D} {\bf 47} 5219
(1993).\\\\
$[19]$ J. D. Barrow and N. J. Nunes, {\it Phys. Rev. D} {\bf 76}
043501 (2007).\\\\

\end{document}